\documentclass[aps, pra, a4paper, 10pt, showpacs, twocolumn]{revtex4-1}
\usepackage{amsmath, amssymb, amsthm, bm, bbm, textcomp}
\usepackage[T1]{fontenc}
\usepackage[latin9]{inputenc}
\usepackage{graphicx,graphics}
\usepackage{geometry}
\usepackage{color}

\newcommand{\one}{\mathbbm{1}}

\newcommand{\C}{\ensuremath{\mathbbm C}}
\newcommand{\be}{\begin{equation}}
\newcommand{\ee}{\end{equation}}
\newcommand{\eea}{\end{eqnarray}}
\newcommand{\bea}{\begin{eqnarray}}

\begin{document}

\title{What we can learn about quantum physics from a single qubit}

\author{Wolfgang D\"ur}
\email{wolfgang.duer@uibk.ac.at} 
\affiliation{Institut f\"ur Theoretische Physik und Institut f\"ur Fachdidaktik, Bereich DINGIM, Universit\"at Innsbruck, Technikerstrasse 25, A-6020 Innsbruck, Austria}

\author{Stefan Heusler}
\email{stefan.heusler@uni-muenster.de}
\affiliation{Institut f\"ur Didaktik der Physik, Universit\"at M\"unster, Wilhelm-Klemm Str. 10, D-48149 M\"unster, Germany}

\date{\today}

\begin{abstract}
We present an approach for teaching quantum physics at high school level based on the simplest quantum system - the single quantum bit (qubit). We show that many central concepts of quantum mechanics, including the superposition principle, the stochastic behavior and state change under measurements as well as the Heisenberg uncertainty principle can be understood using simple mathematics, and can be illustrated using catchy visualizations. We discuss abstract features of a qubit in general, and consider possible physical realizations as well as various applications, e.g. in quantum cryptography.
\end{abstract}

\maketitle

%--------------------------------------------------------------------------------------------------------------------

%--------------------------------------------------------------------------------------------------------------------
\section{Introduction}
%--------------------------------------------------------------------------------------------------------------------

Quantum mechanics is one of the pillars of modern physics, and is still a highly active field of research. Nevertheless, concepts to teach quantum physics at high school level are less developed compared to other areas of physics (see however e.g. \cite{Mu02,Po08,Ca09,Ho07}). On the one hand, this has to do with conceptual difficulties associated with quantum mechanics, which is in conflict with our daily experience and common knowledge \cite{Ba10}. On the other hand, the mathematical formulation of quantum mechanics is rather complex and not suited to high school students without introducing suited simplifications.

Here, we propose an approach based on the simplest quantum mechanical system - the quantum bit (qubit). We show that many of the central concepts of quantum mechanics can be introduced and illustrated with the help of a single qubit. These concepts include the superposition principle and the behavior under measurements, as well as Heisenbergs uncertainty relation. This is done by systematically comparing quantum mechanical states, operations and measurements on the qubit with the case of classical bits. In the case of a single qubit, this can be done with help of simple mathematics involving only two-dimensional vectors, matrix multiplication and scalar products - usually available at high school. What is more, all these concepts and processes can be described using catchy visualizations, which are based on the Bloch sphere representation. These visualizations can be directly used for teaching in class.

Apart from this abstract approach, we discuss various physical realizations of qubits including a spin $1/2$ particle, the polarization degree of freedom of a photon, the position of an atom in a double-well potential, and the electronic degrees of freedom of an ion or atom. The description of states, operations and measurements are developed for all these systems, and experimental realizations are presented. Advantages and disadvantages of the abstract Bloch sphere representation are discussed, where a meaningful treatment in class will certainly involve discussion of the explicit examples for the qubit as well as its  abstract visualization on the Bloch sphere. Finally, we present a number of applications of the developed concepts and features, including a qualitative description of the Heisenberg uncertainty relation, the quantum mechanical no-cloning theorem and quantum cryptography protocols. Interestingly, quantum cryptography is a subject of current research, both theoretically and experimentally, which certainly helps to make the consideration in class more attractive - especially if there is the possibility to visit nearby research laboratories. Following this approach to systems with two and more qubits leads to the discussion of entanglement and its applications in modern quantum information science, with quantum computers and quantum simulators as most prominent examples. This will be discussed in a separate publication. It is also worth mentioning that the Nobel prize in physics 2012 was awarded to S. Haroche and D. Wineland for their contributions to control and manipulate single particles (alias qubits) while maintaining their quantum features \cite{Ha13,Wi13}.

%--------------------------------------------------------------------------------------------------------------------
\section{Classical systems - the classical bit}
%--------------------------------------------------------------------------------------------------------------------

The simplest classical system is a classical bit (binary digit), corresponding to a system with one characteristic property which can have two different states, denoted as $0$ and $1$. The bit is also the smallest unit of classical information, and should be contemplated as abstract object. Various physical realizations of a bit are possible, ranging from a switch that can be on or off, over a voltage with possible values $0V$ or $5V$, to the position of a ball with possible values $x_0$ or $x_1$. Only one characteristic property is considered, and all other features are neglected or fixed to a certain value. In the case of the ball only its position (e.g. on an upper shelf or lower shelf) is important, while its mass, color or size are irrelevant.

\begin{figure}[ht]
\centering
\includegraphics[width=7.5cm]{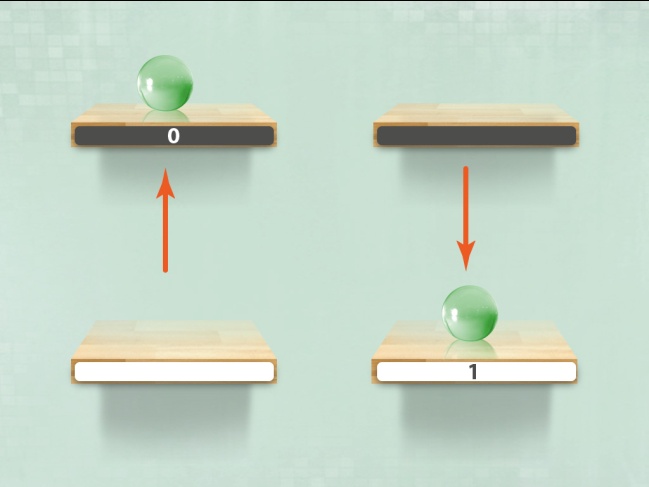}
\caption{Illustration of a classical bit. The two different states $0$ and $1$ are represented by the position of a ball (upper or lower shelf), or the orientation of a vector (up or down).}
\label{FigBit}
\end{figure}

We represent one bit as a vector pointing up (state 0) or down (state 1) - see Fig. \ref{FigBit}. The state of a bit can be changed with help of logical gates, where the $NOT$-gate inverting the bit value $0 \to 1, 1 \to 0$ is the only relevant one in the case of a single bit.

%--------------------------------------------------------------------------------------------------------------------

%--------------------------------------------------------------------------------------------------------------------

%--------------------------------------------------------------------------------------------------------------------
\section{The quantum bit (qubit)}
%--------------------------------------------------------------------------------------------------------------------

We now turn to the simplest quantum mechanical system, the quantum bit or qubit. We consider the description of the system in terms of states, its manipulation by means of operations and measurements, and discuss the resulting properties.

%--------------------------------------------------------------------------------------------------------------------

\subsection{States}
The qubit is the simplest quantum mechanical system and generalizes the classical bit. We consider again a two-level system, i.e. a system with one characteristic property that can have two possible values. All other properties are neglected or assumed to be fixed. In the following, we will develop an abstract mathematical description in parallel with a simple pictorial representation. We believe that for teaching in class, the pictorial approach suffices, and calculations only need to be considered exemplarily. However, we include a complete mathematical description, as we think that this is valuable background information, and is simple enough to be taught in class if desired.

We notate the two states as
\be
|0\rangle =\left(
             \begin{array}{c}
               1 \\
               0 \\
             \end{array}
           \right)
 , |1\rangle=\left(
             \begin{array}{c}
               0 \\
               1 \\
             \end{array}
           \right)
\ee

\begin{figure}[ht]
\centering
\includegraphics[width=7.5cm]{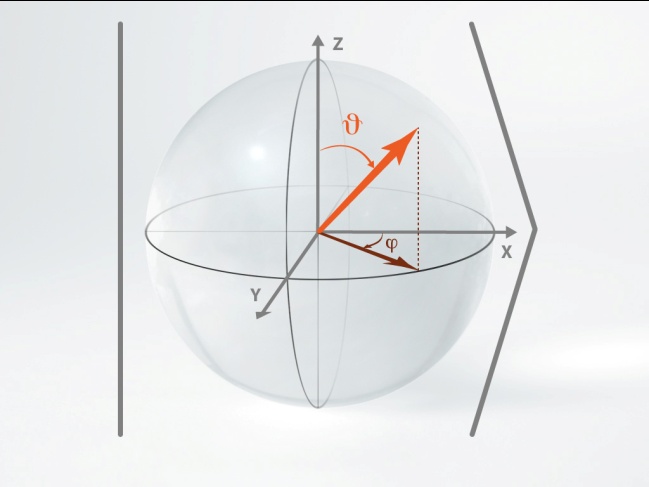}
\caption{Graphical representation of possible states of a single qubit using the Bloch sphere. Quantum mechanical states are described by vectors of length one in the 3-dimensional space, and are characterized by the two angles $\vartheta, \varphi$ of the spherical coordinates.}
\label{FigBlochsphere}
\label{Fig2}
\end{figure}

\begin{figure}[ht]
\centering
\includegraphics[width=7.5cm]{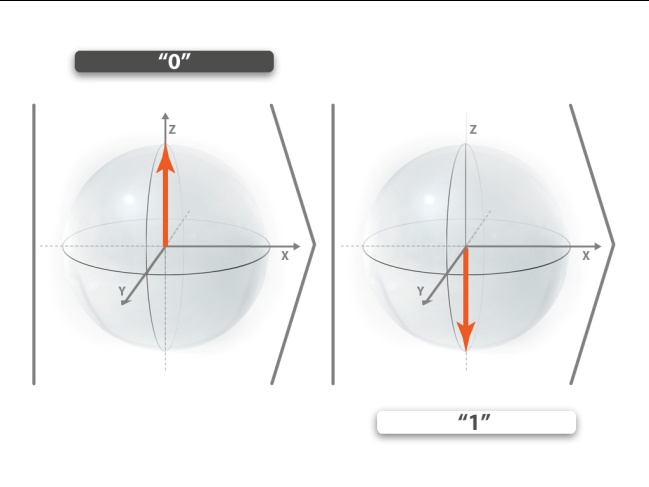}
\caption{Illustration of the two basis states $|0\rangle$ and $|1\rangle$ on the Bloch sphere. Orthogonal states are antiparallel on the Bloch sphere representation.}
\label{FigBlochsphere01}
\end{figure}

A central new feature as compared to a classical bit is the possibility to have superposition states. That is, a qubit can be in an arbitrary superposition of the two basis states $|0\rangle$ and $|1\rangle$. The mathematical description is in terms of a sum of the two basis vectors, weighted by (complex) amplitudes. This corresponds to a 2-dimensional vector with complex coefficients, which is an element of the vector space $\C^2$. Such a superposition might be interpreted as an interference of the two possibilities. The state is described as
\be
\label{EqState}
|\psi\rangle= \alpha|0\rangle + \beta |1\rangle =\left(
             \begin{array}{c}
               \alpha \\
               \beta \\
             \end{array}
           \right).
\ee
Let us consider a second vector, $|\phi\rangle = \gamma |0\rangle +\delta |1\rangle$. Then, the scalar product in $\C^2$ is defined as
$\langle \psi|\Phi\rangle = \alpha^* \gamma + \beta^* \delta$, where $^*$ denotes complex conjugation. In order to allow for a meaningful interpretation of measurements in terms of probabilities --as we will discuss in detail in Sec. \ref{SecMeasurements}-- quantum states need to be normalized, ($\langle \psi|\psi \rangle = |\alpha|^2 + |\beta|^2 = 1)$. In addition, it turns out that a global phase is irrelevant, as all observable quantities do not depend on its value. Hence $\alpha$ in Eq. \ref{EqState} can be chosen real. It follows that $|\psi\rangle$ can be written with help of two real parameters,
\be
\label{EqStateBloch}
|\psi\rangle = \cos{\frac{\vartheta}{2}}|0\rangle + \sin{\frac{\vartheta}{2}}e^{i\varphi}|1\rangle.
\ee
The quantum state of a qubit can be visualized as a vector of length $1$ on the Bloch sphere, see Fig. \ref{FigBlochsphere}. The angles $\vartheta,\varphi$ correspond to the polar and azimuthal angle of spherical coordinates. The two basis states $|0\rangle$ and $|1\rangle$ are represented by $\vartheta = 0$ and $\vartheta=\pi$, respectively, pointing in $+z$ [-$z$] direction (see Fig. \ref{FigBlochsphere01}). Notice that on the Bloch sphere representation, orthogonal vectors are antiparallel.

\begin{figure}[ht]
\centering
\includegraphics[width=7.5cm]{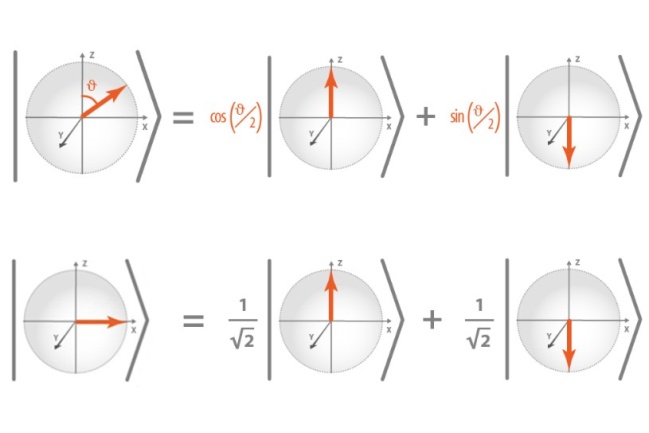}
\caption{Bloch sphere representation of different superposition states.}
\label{FigSuperposition}
\end{figure}

It is now straightforward to depict quantum superposition states. For $\vartheta=\pi/2$ and $\varphi=0$ [$\varphi=\pi$] one obtains e.g. the states
\bea
\label{0x1x}
|0_x\rangle&=& \frac{1}{\sqrt{2}}(|0\rangle + |1\rangle) \\ \nonumber
|1_x\rangle&=& \frac{1}{\sqrt{2}}(|0\rangle - |1\rangle),
\eea
which point in $\pm x$ direction on the Bloch sphere (see Fig. \ref{FigSuperposition}).

It is important to recall that superposition states do not exist for classical systems, and hence they do not have a simple, intuitive meaning. Classical states correspond to a vector pointing in $\pm z$ direction (a ball in upper or lower shelf). Using this picture, a quantum superposition state $|0_x\rangle$ corresponds to a case where the ball is between the two shelves - neither in the upper nor lower shelf, but somewhat in both of them simultaneously. The actual meaning of this will become clearer once we discuss the measurement process in Sec. \ref{SecMeasurements}.

A qubit is an abstract object that may have various physical realizations, which we will discuss in more detail in Sec. \ref{SecRealizations}. It is the basic unit of quantum information, playing a central role in quantum information theory \cite{NiCh}. When introducing quantum states at high school level, it is sufficient to restrict oneself to real coefficients, i.e. $e^{i \varphi}=1$ and $\vartheta \in [0, 2\pi)$. In this way possible mathematical difficulties with complex numbers and scalar products can be avoided. The Bloch-sphere picture reduces to the unit circle, corresponding to unit vectors in the plane that can be parameterized by a single angle. Orthogonal vectors are by convention antiparallel, since we use the angle $\vartheta/2$ in our description of states. The purpose of this convention will become clear when we graphically illustrate the measurement process. Nevertheless, some care and a thorough discussion is required, e.g. when considering qubits realized by the polarization of a single photon.

%--------------------------------------------------------------------------------------------------------------------

\subsection{Operations}
The quantum state of a qubit can be manipulated or evolve in time. This corresponds to a rotation of the state vector on the Bloch sphere, and is mathematically described by a unitary operation, a $2 \times 2$ matrix $U$ from the group $SU(2)$, with $U^\dagger U=U U^\dagger = \one$ where $^\dagger$ denotes complex conjugation and transposition of the matrix. The state after the unitary operation is given by $U |\psi\rangle$. If one uses the bra-ket notation, one can write a unitary operation in the form $U=\sum_{i,j=0}^1 u_{ij}|i\rangle\langle j|$, where $u_{ij}$ are the elements of the matrix.

A rotation with angle $\gamma$ among an arbitrarily oriented axis specified by a normalized vector ${\bf a}=(a_x,a_y,a_z)^T$   is given by
\be
U=\exp(i\gamma \sigma_{\bm a}) = \cos \gamma \one + i \sin \gamma \sigma_{\bm a}
\ee
where
\be
\sigma_{\bm a}={\bm a} \cdot {\bm \sigma}= a_x \sigma_x + a_y \sigma_y + a_z \sigma_z
\ee
and we use the Pauli matrices
\bea
&\sigma_x=\left(
           \begin{array}{cc}
             0 & 1 \\
             1 & 0 \\
           \end{array}
         \right) ,
\sigma_y=\left(
           \begin{array}{cc}
             0 & -i \\
             i & 0 \\
           \end{array}
         \right) ,
\sigma_z=\left(
           \begin{array}{cc}
             1 & 0 \\
             0 & -1 \\
           \end{array}
         \right).
\eea

\begin{figure}[ht]
\centering
\includegraphics[width=7.5cm]{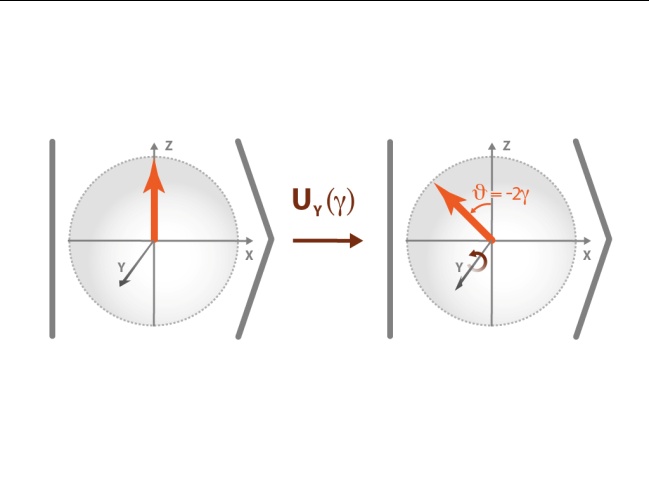}
\caption{A unitary operation acting on a qubit corresponds to a rotation of the state vector with a certain angle among a fixed axes. Here, the rotation $U_y(\gamma)$ acting on the initial state $|0\rangle$ is depicted, corresponding to a rotation with angle $\vartheta=-2\gamma$ among the $y$-axes.}
\label{FigUnitary}
\end{figure}

For instance, a rotation among the $y$-axis with an angle $\vartheta=-2\gamma$ is described by the operation
\be
U_y(\gamma)=\exp(i \gamma \sigma_y) = \left(
           \begin{array}{cc}
             \cos \gamma & \sin \gamma \\
             -\sin \gamma & \cos \gamma \\
           \end{array}
         \right)
\ee
When restricting to real states, this operation is of particular importance. When applied to an initial states $|0\rangle$ we have
\be
U_y(\gamma)|0\rangle = \cos \gamma |0\rangle - \sin \gamma |1\rangle,
\ee
which corresponds to a rotation of the vector among the unit circle, where all possible states can be reached starting from the initial state $|0\rangle$ (see Fig. \ref{FigUnitary}). Similarly, using a rotation among an arbitrary axis, all states on the Bloch sphere can be generated. Notice that any single-qubit rotation can be decomposed into rotations among the $x,y$ and $z$ axis.

%--------------------------------------------------------------------------------------------------------------------

\subsection{Measurements}\label{SecMeasurements}
The presumably most striking and counterintuitive feature of quantum mechanics is the measurement process. All properties of a quantum state can -- in contrast to classical systems-- not simply be determined or read out by a measurement. In case of a qubit, it is only possible to determine one characteristic property with two possible measurement values - for example whether the quantum system is in the state $|0\rangle$ or in the state $|1\rangle$. In general, the measurement result is random, and the the measurement process will change the state of the system.

Mathematically, such a measurement is described by the observable $\sigma_z$, with possible measurement results given by the eigenvalues $+1$ and $-1$ corresponding to the eigenvectors $|0_z\rangle=|0\rangle$ and $|1_z\rangle=|1\rangle$. One of the two possible measurement results $+1,-1$ is obtained, i.e. one (classical) bit of information is learned. Notice that we will sometimes denote the measurement result $+1$ also by $0$ or $|0\rangle$, and similarly the result $-1$ by $1$ or $|1\rangle$ (referring to the associated state).
The physical meaning of the different measurement results depends on the physical realization of the qubit (i.e. the basis states $|0\rangle,|1\rangle$), and will be specified in Sec. \ref{SecRealizations}. If the initial state of the system before the measurement is $|0\rangle$ [$|1\rangle$], the measurement always yields the result $+1$ [$-1$] and the quantum state remains unchanged. In this case, the qubit behaves like a classical bit.

However, if the state of the qubit is initially given by the superposition state $|\psi\rangle = \alpha |0\rangle + \beta |1\rangle$, upon measurement, one obtains a random, unpredictable outcome of  $+1$ (corresponding to the state $|0\rangle$ and labeled 0) or $-1$  (corresponding to $|1\rangle$, labeled 1). But quantum randomness is not completely unpredictable: On the one hand, for certain states the measurement result is deterministic. For the $z$-measurement, this is the case for states $|0\rangle$ and $|1\rangle$. On the other hand, even though we are not able to make any predictions on an individual event, the statistical behavior for repetitions of the experiment (i.e. state preparation followed by measurement) can be predicted. The probability to obtain a certain measurement result is given by
\bea
p_0&=&\langle \psi|0\rangle\langle 0 |\psi\rangle=|\langle 0|\psi\rangle|^2 =|\alpha|^2, \\ \nonumber
p_1&=&\langle \psi|1\rangle\langle 1 |\psi\rangle=|\langle 1|\psi\rangle|^2 =|\beta|^2.
\eea
Notice that the probabilistic behavior of an individual measurement is not due to limited information about the state as e.g. in classical statistical mechanics, but is an intrinsic feature of the quantum mechanical description. Even if we knew the initial state,  intrinsic randomness for individual events upon measurement emerges. Thus, quantum mechanics can only provide statistical predictions, expressed in terms of probabilities for certain events when considering multiple repetitions of the experiment.

After the measurement, the state of the qubit has changed and is no longer given by the initial superposition state. In particular, if the measurement result is $+1$, then the state of the system after the measurement is $|0\rangle$. Similarly, if the measurement result is $-1$, the state after the measurement is given by $|1\rangle$. Again, this is in sharp contrast to the behavior of classical systems.

%--------------------------------------------------------------------------------------------------------------------

\subsection{Measurements in a rotated basis}
It is also possible to measure alternative properties of the same initial state, e.g. whether it is in the state $|0_x\rangle$ or $|1_x\rangle$ (Eq. \ref{0x1x}). Mathematically, this corresponds to the measurement of the observable $\sigma_x$ with eigenvectors $|0_x\rangle,|1_x\rangle$ and eigenvalues $+1,-1$. If the initial state is $|0_x\rangle$, then such a measurement yields the outcome $+1$ with probability 1, while now an initial state $|0\rangle$ leads to a probabilistic outcome. Again, the state of the system after the measurement is given by $|0_x\rangle$ for outcome $+1$, and $|1_x\rangle$ for outcome $-1$.

In general, a measurement in an arbitrary basis is possible. The measurement is specified by two orthogonal vectors $|0_{\bm a}\rangle,|1_{\bm a}\rangle$, the eigenvectors with eigenvalues $+1$ and $-1$ of an observable $A$. The observable is defined as
\be
A=(+1) |0_{\bm a}\rangle \langle 0_{\bm a}| + (-1)|1_{\bm a}\rangle \langle 1_{\bm a}|,
\ee
and we have $A|0_{\bm a}\rangle = |0_{\bm a}\rangle$, $A|1_{\bm a}\rangle = -|1_{\bm a}\rangle$. The probability to obtain the outcome $+1$ (corresponding to $|0_{\bm a}\rangle$) or $-1$ (corresponding to $|1_{\bm a}\rangle$) is given by the scalar product with the initial state,
\bea
\label{Eqprob}
p_0=|\langle 0_{\bm a}|\psi\rangle|^2 &,& p_1=|\langle 1_{\bm a}|\psi\rangle|^2.
\eea
From these formulaes it is immediately clear that states of the form $e^{i\gamma}|\psi\rangle$ are physically equivalent to the state $|\psi\rangle$, i.e. a global phase is unimportant for measurements.
The state after the measurement either is given by $|0_{\bm a}\rangle$, or by $|1_{\bm a}\rangle$.

We emphasize that since the measurement changes the initial state to an eigenstate of the measured observable, all information about the initial state are erased upon measurement. Therefore, subsequent measurements cannot reveal additional information about the initial state. The total information gain for one qubit is hence restricted to one classical bit of information.

%--------------------------------------------------------------------------------------------------------------------

\subsection{Visualization of the measurement process on the Bloch sphere}
The measurement process can be visualized using the Bloch sphere picture. Notice that two orthogonal states $|0_{\bm a}\rangle , |1_{\bm a}\rangle$ lie on opposite points of the Bloch sphere, and together specify a direction ${\bm a}$ in space \footnote{To be precise, the direction is given by $\lambda {\bm a}$}. For instance, the two states $|0\rangle, |1\rangle$ lie on two opposite poles of the Bloch sphere, pointing in $\pm z$ direction and specify the $z$-axis.

\begin{figure}[ht]
\centering
\includegraphics[width=7.5cm]{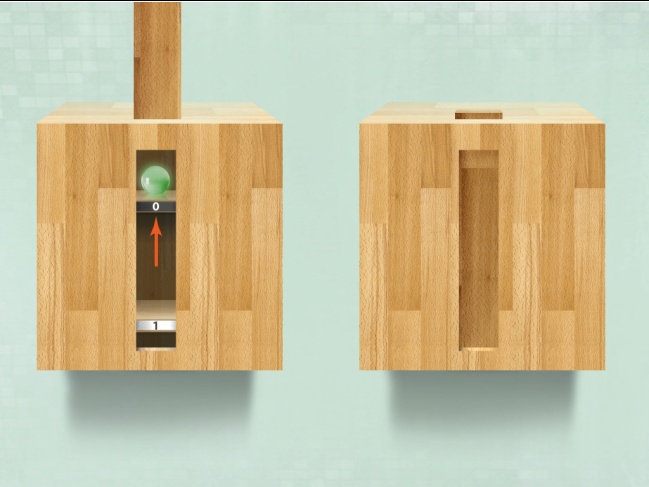}
\caption{Illustration of a $z$-measurement (observable $\sigma_z$), corresponding to orientation of the slit in $z$-direction. The measurement determines whether the state vector is oriented in $+z$ or $-z$ direction (measurement result $|0\rangle$ or $|1\rangle$).}
\label{Fig6}
\end{figure}

\begin{figure}[ht]
\centering
\includegraphics[width=7.5cm]{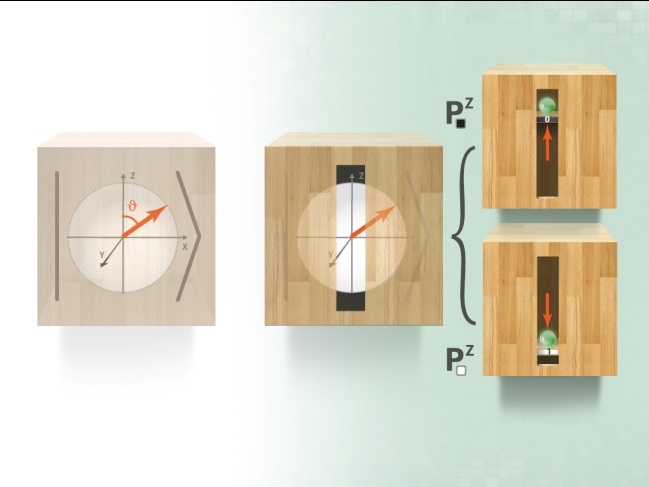}
\caption{Illustration of a $z$-measurement (observable $\sigma_z$) performed on a qubit in state $|\psi\rangle = \cos{\frac{\vartheta}{2}}|0\rangle + \sin{\frac{\vartheta}{2}}|1\rangle$. The state vector is not oriented in slit direction, and hence the measurement process enforces a rotation of the vector in positive or negative $z$-direction. This leads to a random measurement result $|0\rangle$ or $|1\rangle$, with probability $p_0=\cos^2{\frac{\vartheta}{2}}$ and $p_1=\sin^2{\frac{\vartheta}{2}}$, and a change of the state vector after the measurement.}
\label{Fig7}
\end{figure}

We propose a slit oriented in some direction ${\bm a}$ for the illustration of the measurement axis, corresponding to a measurement in the basis $\{|0_{\bm a}\rangle,|1_{\bm a}\rangle\}$. The standard measurement of $\sigma_z$ corresponds to a slit in z-direction. If the initial state vector already points into the direction of the slit, it can simply pass through. We obtain a deterministic outcome, and the state vector does not change (see Fig. \ref{Fig6}). If, however, the state vector is not parallel to the slit, it needs to be pushed through the slit and hence change orientation, pointing either in $+z$ or $-z$ direction afterwards. Hence the state of the system has changed due to the measurement process. If the state is already close to $+z$ direction, it is likely that it will point in $+z$ direction afterwards, i.e. the probability to obtain the outcome $+1$ is close to 1. In turn, it is unlikely that the state is flipped all the way down to $-z$ direction, and hence the probability to obtain the outcome $-1$ is small (see Fig. \ref{Fig7}). For a state vector pointing e.g. in $x$-direction ($|\psi\rangle=|0_x\rangle$), it is equally likely to point in $+z$ or $-z$ direction after the measurement. In general, the angle between the state vector and the orientation of the slit determines the measurement probabilities, which is explicitly expressed in Eq. \ref{Eqprob}.

\begin{figure}[ht]
\centering
\includegraphics[width=7.5cm]{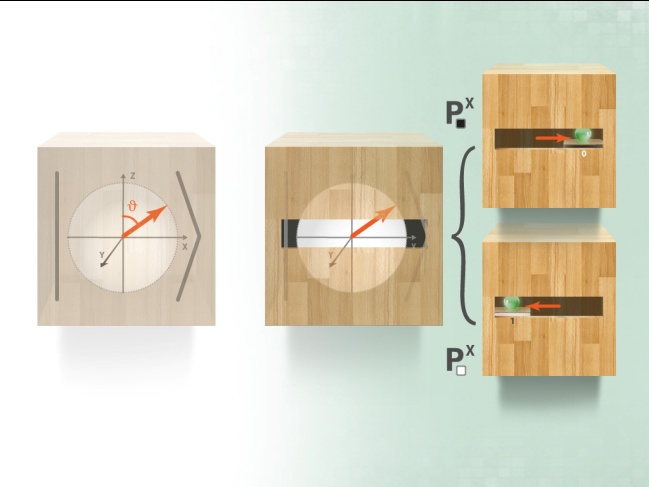}
\caption{Illustration of a $x$-measurement (observable $\sigma_x$) performed on a qubit in state $|\psi\rangle = \cos{\frac{\vartheta}{2}}|0\rangle + \sin{\frac{\vartheta}{2}}|1\rangle$. The slit is oriented in $x$-direction. The measurement process enforces a rotation of the vector in positive or negative $x$-direction. This leads to a random measurement result $|0_x\rangle$ or $|1_x\rangle$, with probability $p_0=1/2+\cos{\frac{\vartheta}{2}}\sin{\frac{\vartheta}{2}}$ and $p_0=1/2-\cos{\frac{\vartheta}{2}}\sin{\frac{\vartheta}{2}}$, and a change of the state vector after the measurement.}
\label{Fig8}
\end{figure}

A slit pointing in $x$-direction corresponds to a measurement of the observable $\sigma_x$, leaving the system in either the state $|0_x\rangle$ (measurement result $+1$) or $|1_x\rangle$ (measurement result $-1$) after the measurement (see Fig. \ref{Fig8}). If the initial state was $|0_x\rangle$, i.e. a vector pointing in $+x$ direction, it will simply pass through the slit, leading to the deterministic outcome $+1$, leaving the state unchanged.

This visualization makes it easy to grasp that it is impossible to determine the initial state of the system with a single measurement. The slit limits the view of the state vector to one direction on the Bloch sphere. In addition, probabilistic measurement results as well as the state change due to the measurement are made plausible. However, we should emphasize that the slit just illustrates the measurement process, and does not have any direct physical meaning.

%--------------------------------------------------------------------------------------------------------------------

\subsection{Classical mixtures vs. quantum superpositions}
\begin{figure}[ht]
\centering
\includegraphics[width=7.5cm]{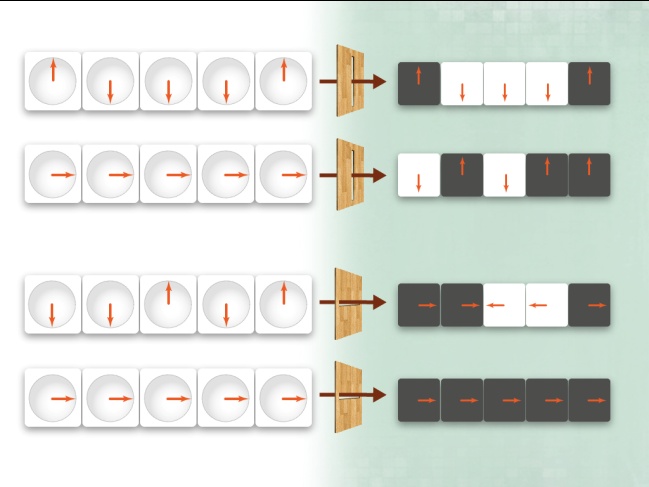}
\caption{Illustration of the difference between a mixed state (ensemble of random bits in state $|0\rangle$ or $|1\rangle$) and a pure state (all qubits in state $|0_x\rangle=1/\sqrt{2}(|0\rangle + |1\rangle)$). The two situations are indistinguishable by $z$-measurements. In both cases one obtains random outcomes (see line 1 and 3). A $x$-measurement, however, leads to random measurement results for the mixed state (line 3), while for the pure states, always the same measurement result $|0_x\rangle$ is found.}
\label{Fig9}
\end{figure}

We now discuss the difference between classical mixtures and a quantum mechanical ensembles of superposition states. Also classical system can show a stochastic behavior under measurements, however, the underlying reason is completely different compared to the quantum mechanical case. To illustrate this, let us consider an ensemble of $N$ classical bits, where each bit has a fixed but random value $0$ or $1$. We are hence dealing with a situation where each bit has a fixed value $0$ or $1$, however we do not know this value and hence call it random. When performing measurements on these $N$ bits, we will find random outcomes. The statistics of the measurement is such that we will find the value $0$ or $1$ in approximately $N/2$ cases. Such a situation can be described using a probability distribution for the possible bit values, which expresses our lack of knowledge about the situation in question. Such a description is e.g. applied in statistical mechanics. One obtains exactly the same measurement results when considering an ensemble of random quantum bits, where each of the qubits is in either the state $|0\rangle$ or $|1\rangle$ with equal probability (Fig. \ref{Fig9}, first line). This is a so-called mixed state, described by a density matrix $\rho=q_0 |0\rangle\langle 0| +q_1 |1\rangle \langle 1|$. Mixed states can either be visualized as shown in the figure, or by just one Bloch sphere with a single vector of length $\leq 1$. In our example with $q_0=q_1=1/2$, where the probabilities are equal for both states, the length of the effective Bloch vector is $0$, corresponding to a completely mixed state. This means that measurements in {\em any} basis lead to completely random outcomes, which can be seen by noting that $\one = (|0_{\bm a}\rangle\langle 0_{\bm a}| + |1_{\bm a}\rangle \langle 1_{\bm a}|)/2$ for any ${\bm a}$ .

Let us now compare this with a situation where we have $N$ qubits, each of them in the same quantum superposition state $|0_x\rangle =1/\sqrt{2}(|0\rangle + |1\rangle)$ (Fig. \ref{Fig9}, second line). If we perform a $z$-measurement on each of the qubits, the result will be random, where we find the results $|0\rangle$ and $|1\rangle$ with probability $1/2$ each. On average, we obtain the results $|0\rangle$ or $|1\rangle$ in approximately $N/2$ cases, analogous to the classical mixture and to the quantum mixture discussed above. In fact, these ensembles are indistinguishable by $z$-measurements.

However, if we perform a $x$-measurement, we obtain again a completely random outcome for each measurement for the mixed quantum states (Fig. \ref{Fig9}, third line). The reason is that for both possible initial states $|0\rangle$ and $|1\rangle$, the probability to obtain the results $+1$ and $-1$ when performing an $x$-measurement is given by $1/2$, i.e. the outcomes are equally probable in both cases. This can be easily visualized on the Bloch sphere representation, where the slit is oriented along the $x$-direction, and the state vector of both initial states $|0\rangle$ and $|1\rangle$ is perpendicular to the slit orientation. In contrast, for the pure state $|0_x\rangle$, we have a deterministic outcome, i.e. we obtain with probability $1$ the result $|0_x\rangle$ ($+1$) (Fig. \ref{Fig9}, fourth line). The two situations are hence different: while for the classical mixture, we lack information about the ensemble, leading to random outcomes for all measurements, in the second situation the system is in a unique pure state. It posses the feature of pointing in $+x$ direction (which can be checked by a $x$-measurement). However, one can not assign the property to point in $+z$ or $-z$ direction to the quantum state before the measurement - when measuring this property, one obtains a random outcome. We also remark that for classical bits, an $x$-measurement is not defined and hence impossible, as classical bits unlike quantum bits do not show interference properties.

The proposed visualizations of qubits, operations on qubits, and of the measurement process can be used to illustrate three central features of quantum mechanics:
\begin{itemize}
\item The superposition principle
\item Stochastic behavior under measurements
\item State change due to measurement
\end{itemize}
In addition to the qualitative pictorial description with vectors and slits on the Bloch sphere that helps to visualize these concepts and processes, a simple quantitative description that only involves scalar products and two-dimensional vectors can be introduced.

%--------------------------------------------------------------------------------------------------------------------
%--------------------------------------------------------------------------------------------------------------------
\section{Physical realizations of a qubit}\label{SecRealizations}
%--------------------------------------------------------------------------------------------------------------------
%--------------------------------------------------------------------------------------------------------------------

A qubit is an abstract object, and we have treated it this way up to now to emphasize the quantum mechanical features, as compared to a classical bit. Many possible physical realizations of a qubit exist, where we discuss a number of examples in the following. For a treatment in class, we suggest to illustrate the abstract properties of a qubit with help of several physical realizations. Notice that the different systems allow for a proper illustration of certain aspects (states, measurement, operations), however none of the systems is free from possible misconceptions or potential problems, which might occur in classroom.

Qubits have been realized in the laboratory with many different systems, thereby experimentally confirming the strange quantum mechanical properties and the behavior under measurements. Some of these systems show an extraordinary controllability, and states, operations and measurements can be implemented with almost unit fidelity. In particular, experiments with single photons, atoms or ions are performed by various groups world-wide, where the capability to manipulate light and matter at a microscopic level is clearly demonstrated \cite{}.

%--------------------------------------------------------------------------------------------------------------------

\subsection{Polarization degree of freedom of a photon}
\label{Photon}
\begin{figure}[ht]
\centering
\includegraphics[width=7.5cm]{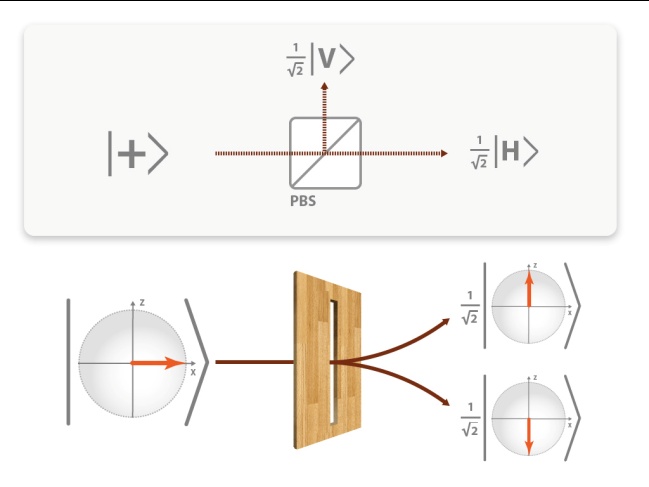}
\caption{Illustration of the measurement of the polarization degree of freedom of a single photon. The polarizing beam splitter leads to a transmission of horizontally polarized photons, while vertically polarized photons are reflected by the beam splitter. Photons are detected at both branches with help of single-photon detectors, where only one of the two detectors will register a photon. This corresponds to a $z$-measurement. i.e. measurement of the observable $\sigma_z$.}
\label{Fig10}
\end{figure}

One possibility to realize a qubit are single photons, where e.g. the polarization degree of freedom can be used. The state $|0\rangle = |H\rangle$ is given by the horizontal polarization of the photon, while $|1\rangle =|V\rangle$ is given by the vertical polarization. The superposition state $|0_x\rangle$ corresponds to a polarization of $45^\circ$. The manipulation of the photonic state is achieved by wave plates ($\lambda$-plates), leading to a rotation of the polarization state vector. Measurements can be performed with help of a polarizing beam splitter, where a horizontally polarized photon is transmitted, while a vertically polarized photon is deflected by $90^\circ$ (see Fig. \ref{Fig10}). With help of single-photon detectors at the two outputs of the polarizing beam splitter, one can distinguish between the two states $|H\rangle$ and $|V\rangle$. The whole procedure corresponds to a $z$-measurement (measurement of the observable $\sigma_z$). Measurements in a different basis may be realized by first rotating the polarization appropriately by means of wave plates, followed by a $\sigma_z$ measurement.

If a $45^\circ$ polarized photon $|0_x\rangle=1/\sqrt{2}(|H\rangle + |V\rangle)$ passes through the polarizing beam splitter, only one of the two detectors will click and register the photon. The photon will be registered with probability $1/2$ by detector 1 after being transmitted, and with probability $1/2$ by detector 2 after reflection. Notice that only due to the measurement process (i.e. the detectors), the superposition state is destroyed. The superposition is $not$ destroyed due to the polarizing beam splitter. Here, the wave function of the photon is just divided into two spacially disctinct parts. This can be nicely illustrated by adding two mirrors and an additional beam splitter to the set-up to form an interferometer. There, superposition and interference play a central role, and in fact the superposition state is restored at the output of the interferometer.

When using polarization states to realize a qubit, one picks up a well-known concept from classical optics the students are familiar with. On the one hand, this helps to visualize not only basis states $|H\rangle$ and $|V\rangle$, but also superposition states such $|0_x\rangle$ corresponding to $45^\circ$ polarized light or a $45^\circ$ polarized photon. Furthermore, operations and measurements are intuitive and transparent. On the other hand, using the classical concept of polarization bears the danger that new quantum mechanical features are not recognized and viewed as properties that also classical objects (in this case a classical electromagnetic wave) possess. In the classical case, however, one deals not with single photons but with an electromagnetic wave, where the typical quantum mechanical features such us the behavior under measurements or the quantization of the polarization degrees of freedom are absent. Classical light described by an electromagnetic wave is partly transmitted and partly deflected at a polarizing beam splitter, where only the amplitude of each of the beams is decreased as compared to the incoming beam. Only due to the quantization of light and the descriptions in terms of single photons it is legitimate to talk about a quantum object - a qubit, where one photon is  -  before the measurement - transmitted and reflected at the same time.

\begin{figure}[ht]
\centering
\includegraphics[width=7.5cm]{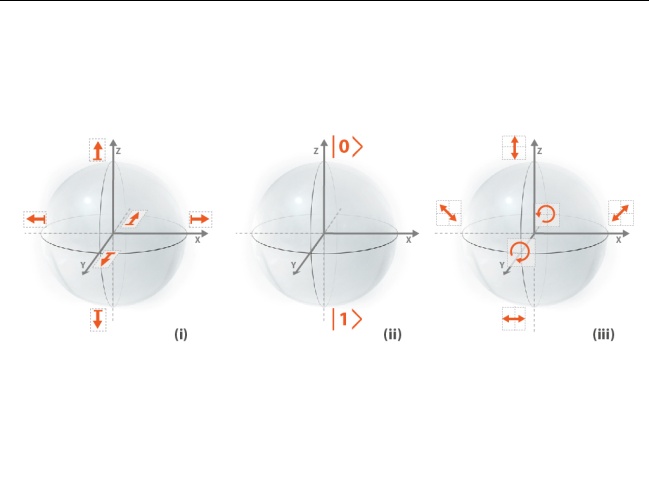}
\caption{Comparison of Bloch sphere representation for spin and photon. While the orientation of the spin is given by the orientation of the Bloch vector, orthogonal polarization states such as $|H\rangle$ and $|V\rangle$ are antiparallel on the Bloch sphere representation. A $45^\circ$ polarized photons corresponds to a Bloch vector in $\pm x$ direction, while circular polarized photons are described by a Bloch vector in $\pm y$ direction.}
\label{Fig11}
\end{figure}

In addition, the Bloch sphere representation can be misleading and might easily lead to confusion when not properly discussed. On the Bloch sphere representation, orthogonal vectors are antiparallel (e.g. horizontal and vertical polarization correspond to a Bloch vector in positive or negative $z$-direction), while polarization vectors corresponding to orthogonal states (e.g. horizontal and vertical polarization) are orthogonal in the laboratory, enclosing an angle of $90^\circ$. The origin of these difficulties lies in the definition of the angle $\vartheta/2$ in the abstract Bloch sphere representation, which is perfectly suited to describe other systems such as spins, and also allows for a simple pictorial representation of the measurement process with help of slits. Notice that all possible polarization states of a single photon can be described on the Bloch sphere representation, including e.g. circular polarized photons described by $|0_y\rangle = \frac{1}{\sqrt{2}}(|0\rangle \pm i |1\rangle)$ pointing in $\pm y$ direction on the Bloch sphere (see Fig. \ref{Fig11}).

 Experiments with single photons are very advanced. Production, manipulation and transmission of single photons can be achieved with high accuracy, thereby nicely illustrating quantum mechanical features. The detection of single photons has also been significantly improved in recent years, however non-unit efficiency and dark counts still pose a problem in certain experiments, e.g. for loophole free tests to violate Bell inequalities. In quantum communication and quantum cryptography, the realization of qubits with the help of photons plays an important role, because they are perfectly suited for transmission of quantum information via optical fibers or even free space. Experiments implementing quantum communication or establishing entangled photons over distances of more than 100km have been performed, and proposals for satellite-based quantum communication together with first proof of principle experiments are being pursued \cite{ExpPhoton_longRange}. We remark that apart from using the polarization degree of freedom, there are a number of different ways to realize a qubit with help of a single photon.  This includes e.g. an encoding corresponding to time delays (time-bin photons), or the usage of different spatial modes (where a photon in the left mode corresponds to state $|0\rangle$, and in the right mode to state $|1\rangle$). Also the usage of microwave photons trapped in a cavity (cavity quantum electrodynamics) should be mentioned \cite{Ra01,Ha13}.  For the treatment in class, nicely elaborated virtual experiments can be used (see \cite{Qlab})

%--------------------------------------------------------------------------------------------------------------------

\subsection{Spin of a particle}
\label{Spin}
The spin of a particle is the textbook example for a discrete quantum mechanical system, which is also of particular interest from a historical point of view. The spin is a quantum mechanical property without any classical counterpart, and may be interpreted as intrinsic angular momentum of the particle. A qubit is associated with a spin $1/2$ particle, where $|0\rangle =|\uparrow\rangle$ and $|1\rangle = |\downarrow\rangle$ denote the two basis states of the spin, corresponding to an orientation of the spin in positive or negative $z$-direction respectively. Superposition states correspond to an orientation of the spin in a different spatial direction, e.g. the state $|0_x\rangle$ (Eq. \ref{0x1x}) is oriented in $+x$ direction. In fact, the spatial orientation of the spin coincides with the abstract Bloch sphere representation, which is hence especially suited to illustrate it (see Fig. \ref{Fig11}).

\begin{figure}[ht]
\centering
\includegraphics[width=7.5cm]{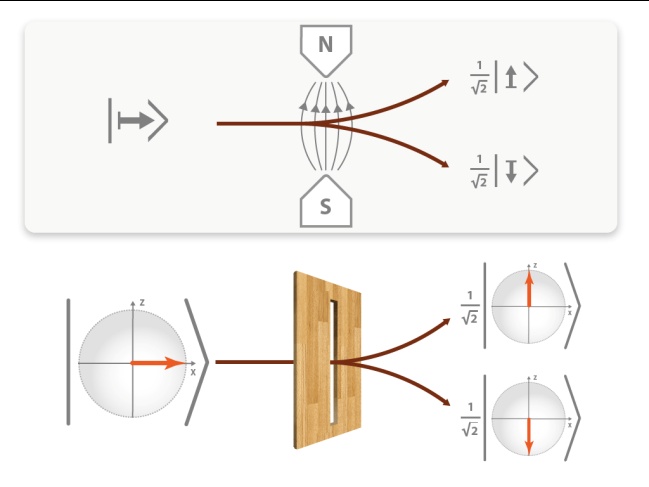}
\caption{Illustration of the measurement process for a spin with a Stern-Gerlach apparatus. The orientation of the inhomogeneous magnetic field determines the measurement direction (slot). Depicted is a measurement in $z$-direction. Coupling of the spin to the inhomogeneous magnetic filed leads to a discrete displacement of the particle, corresponding to the two measurement results $+$ (spin up) or $-1$ (spin down). }
\label{Fig12}
\end{figure}

The state of a spin can be changed by applying an external, homogeneous magnetic field in some spatial direction. The spin (or the associated magnetic moment) performs a precision about the magnetic field axes with the so-called Larmor frequency. The rotation angle can be adjusted by varying the time for which the magnetic field is applied.
A measurement can be performed with help of a Stern-Gerlach apparatus (see Fig. \ref{Fig12}). The particle passes through an inhomogeneous magnetic field, leading to a discrete displacement depending on the orientation of the spin. The magnetic field needs to be inhomogeneous such that the spin does not perform a precision, but becomes oriented in field direction. Experimentally, only discrete deflections are found since the spin can only take discrete values. Historically, with such an experiment (performed with silver atoms) the quantization of the spin was demonstrated for the first time. The measurement direction or basis is determined by the orientation of the inhomogeneous magnetic field, where an orientation in ${\bm a}$ direction corresponds to the measurement of the observable $\sigma_{\bm a}$, i.e. a slot in ${\bm a}$ direction on the Bloch sphere.

Therefore, states, operations and measurements for a spin can be nicely interpreted and illustrated on the Bloch sphere representation. A problem for explanations in class might be the missing interpretation of the involved concept - there is no classical analogy for a spin, apart from the interpretation as intrinsic angular momentum which is however hard to contemplate for a point particle.

Experimentally, single electrons are stored in quantum dots with help of electric fields, and the electron spin is used to represent a qubit. Based on this principle, state preparation, gates and measurements have been experimentally demonstrated, although the achieved qualitiy is a bit less as compared to photons or ions. Proposals to realize a large-scale quantum information processing device based on such a set-up exist \cite{Ta05} . Due to their long coherence times, nuclear spins have been discussed and already been used to store quantum information.

%--------------------------------------------------------------------------------------------------------------------

\subsection{Position of an atom}
\begin{figure}[ht]
\centering
\includegraphics[width=7.5cm]{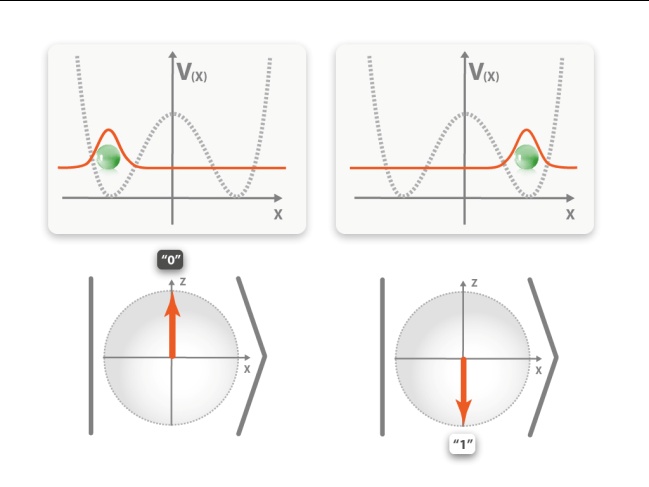}
\caption{Realization of a qubit using spatial degrees of freedom of a single atom. The state $|0\rangle$ corresponds to the localization of the atom in the left well of the double well potential, while the state $|1\rangle$ corresponds to the localization in the right well.}
\label{Fig13}
\end{figure}

\begin{figure}[ht]
\centering
\includegraphics[width=7.5cm]{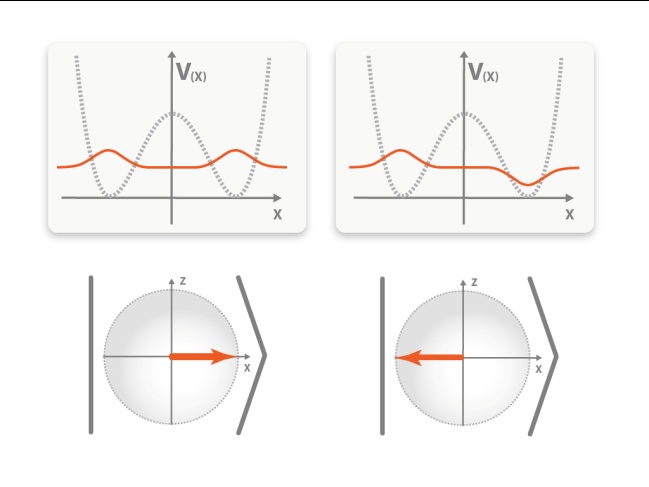}
\caption{Superposition states of an atom in a double well potential. The state $|0_x\rangle$ corresponds to a symmetric superposition, while the state $|1_x\rangle$ corresponds to the antisymmetric superposition.}
\label{Fig14}
\end{figure}

Another possible realization of a qubit is based on the usage of spatial degrees of freedom, where only two positions $x_0,x_1$ are considered. Classically, this corresponds to a ball that is either in the upper or lower shelf. For an atom, this means that the atom is either at position $x_0$ corresponding to state $|0\rangle$, or at position $x_1$ corresponding to state $|1\rangle$ \cite{Mo03} (see Fig. \ref{Fig13}). Superposition states of an atom are particularly counter-intuitive and difficult to arrange with our classical everyday experience. Particles can only be at one position - and not at two positions simultaneously. Despite of this, it is possible to prepare such superposition states experimentally, where a single atom is trapped in a double well potential (see Fig. \ref{Fig14}). The manipulation of the potential barrier can be used to control a coherent tunneling process and thus manipulate the quantum state of the atom. The $z$-measurement corresponds to a simple position measurement, i.e. to check in which of the two wells the particle is located. In this picture, the strange properties of quantum superposition are highlighted -- namely the existence of superposition states  where the position of the atom is not specified. This is also the underlying principle of the famous Schr\"odinger cat, illustrating these counterintuitive features of quantum mechanics which are in conflict with our everyday experience.

In this example, a macroscopic object (e.g. a cat) is in a superposition state $(|0\rangle|{\rm cat}_{\rm alive}\rangle + |1\rangle|{\rm cat}_{\rm dead}\rangle)/\sqrt{2}$. The state of the cat depends on the state of a radioactive atom (first system). If the atom is not decayed (state $|0\rangle$), the cat is alive, while the decay of the atom (state $|1\rangle$) triggers some poison leading to the death of the cat. The question if the cat is dead or alive as long as nobody observes the system cannot be satisfactory answered with help of these classical terms, which only allow for a ``either - or'', but not for an ``as well as'' or ``both at the same time''. The cat is in a superposition state where this property (dead or alive) is not specified.

Only the measurement changes the state of the cat, and after the measurement the cat is either dead or alive. From a didactical point of view, there are a number of difficulties with this thought experiment. On the one hand, we no longer deal with just a single quantum qubit, but with composite, high dimensional systems. On the other hand, the term observation is in its everyday usage linked with human senses and consciousness. In quantum mechanics, however, an observation has in general no relation to human sensation or consciousness, but refers to an interaction of a quantum system with a measurement apparatus that leads to a destruction of the quantum superposition. This process is visualized in Figs. \ref{Fig7},\ref{Fig8}.

For a macroscopic object, such superposition states are hard to prepare and difficult to maintain, but according to the laws of quantum mechanics they are in principle possible \footnote{Note that modifications of quantum mechanics have been suggested that make such macroscopic superposition states instable and hence impossible to prepare or observe, e.g. gravitational collapse models by Penrose or the GRW model. However they are not supported by experimental data so far.}. In modern experiments, the limits of quantum mechanics are being investigated, where one aims at producing superposition states of large quantum mechanical objects, including e.g. a small mirror or a large number of photons, atoms or ions \cite{IonsExp,Macroscopic,Ra01,Ha13,Wi13} (see also \cite{Fr12} for a recent theoretical discussion on measures for the degree of macroscopicity). In addition, wave properties of massive objects such as large molecules with several thousand mass units are being tested by transmitting them through a double slit or grid, and observing an interference pattern \cite{Interference}.

Here, we should warn against using quantum mechanical terminology together with metaphysical terms such as live, death or consciousness. This leads --at least at the novice level-- to confusion and misunderstandings. To avoid such difficulties, it is better to talk about quantum superposition states of macroscopic objects , where e.g. an object consisting of a large number of atoms is either located at position $x_0$ or $x_1$. This already stretches our imagination sufficiently, and better reflects current experiments and experimental proposals.

%--------------------------------------------------------------------------------------------------------------------

\subsection{Electronic degrees of freedom of an ion or atom}

Another possibility for the realization of a qubit based on single atoms or ions are two internal (electronic) states. All other degrees of freedom of the atom are frozen or being neglected. In particular, the atom is cooled via laser cooling to the motional ground state. Ions are trapped in a Paul trap with help of electromagnetic fields, where a rotating saddle potential enforces a dynamical trapping in all three spatial directions. The quantum state of the ion is manipulated by laser pulses of certain frequency and duration, which couples two electronic states either directly or via a Raman transition. The measurement process is realized by an additional, meta-stable auxiliary level. A laser pulse couples the internal state $|0\rangle$ to the auxiliary level, thereby exciting the ion. The auxiliary state decays back to the state $|0\rangle$, where a photon is emitted and subsequently detected. The detection of the photon corresponds to the measurement result $+1$ and the ion is left in state $|0\rangle$, while a measurement result $-1$ is found if no photon is detected, leaving the ion in state $|1\rangle$. To achieve a high detection efficiency, light is scattered on this transition until several photons have been detected, e.g. by a CCD camera.

\begin{figure}[ht]
\centering
\includegraphics[width=7.5cm]{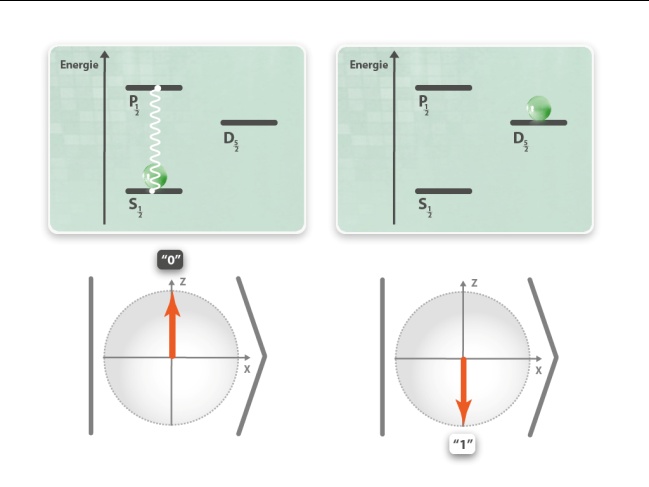}
\caption{Realization of a qubit using electronic states of an ion or atom (atomic level scheme). The state $|0\rangle$ corresponds to the occupation of the $S_{1/2}$ level, while the state $|1\rangle$ corresponds to the occupation of the $D_{5/2}$ level. A measurement is realized by scattering laser light on the $S_{1/2}-P_{1/2}$ transition.}
\label{Fig15}
\end{figure}

For a better illustration, it may be useful to refer to known atomic models such as the Rutherford model. In this model, the state $|0\rangle$ corresponds to an electron at an inner orbit (with small radius), while the state $|1\rangle$ corresponds to an excited state at an outer orbit (with larger radius). The actual experimental realization makes use of different atomic levels -- for instance experiments with $^{40}{\rm Ca}^+$-ions the levels $S_{1/2}$ ($m=-1/2$) and $D_{5/2}$ ($m=-1/2$) are used to encode states, and the $P_{1/2}$ level is used as auxiliary level for measurements, where light is scattered on the $S_{1/2}-P_{1/2}$ transition (see Fig. \ref{Fig15}). Other experiments, e.g. with neutral atoms, make use of hyperfine levels. Experiments with single atoms and ions are very advanced, and extraordinary control at the single atom level has been achieved \cite{ions,Wi13}. It is possible to control and manipulate the states of the atoms with a fidelity of $99.99\%$, and also measurements can be performed with similar accuracy. In fact, cold atoms and ions are a promising platform for quantum information processing, where also entangled states of up to 14 ions have been experimentally created \cite{IonsExp,ions,Wi13}, or a large number of atoms have been stored and interfere in a controlled manner in optical lattices \cite{OpticalLattice}.

%--------------------------------------------------------------------------------------------------------------------
%--------------------------------------------------------------------------------------------------------------------
\section{Applications}
%--------------------------------------------------------------------------------------------------------------------
%--------------------------------------------------------------------------------------------------------------------
In the following, we will discuss a number of applications to illustrate the principles discussed in the previous sections. Our focus of interest lies at the behavior of the system under measurements. We will consider fundamental aspects, including the possibility to determine or copy an unknown quantum state, as well as a (qualitative) discussion of Heisenbergs uncertainty relation. Finally, we describe an application in the context of quantum cryptography, i.e. the secure transmission of secret messages.

%--------------------------------------------------------------------------------------------------------------------

\subsection{State tomography}
Due to the behavior of a quantum system under measurements, it is impossible to determine the state of an (unknown) qubit completely. Any measurement does only yield one classical bit of information, the answer to a single yes/no question. The state of the system is changed due to the measurement, and hence no further information about the initial state of the system can be obtained by subsequent measurements. To illustrate this, we consider an unknown qubit in state (see Fig. \ref{Fig2})
\be
|\psi\rangle =\cos \frac{\vartheta}{2}|0\rangle + \sin \frac{\vartheta}{2} e^{i \varphi} |1\rangle.
\ee
A $z$-measurement yields with probability $p_0=\cos^2\frac{\vartheta}{2}$ the result $+1$, and with probability $p_1=\sin^2\frac{\vartheta}{2}$ the result $-1$. After the measurement, the state of the qubit is given by $|0\rangle$ or $|1\rangle$ respectively, and does no longer contain any information on the parameters $\vartheta$ and $\varphi$ (see Fig. \ref{Fig7}). In addition, the measurement result does not provide us with the full knowledge about these parameters: On the one hand, the measurement result does not depend on the parameter $\varphi$, so no information about this parameter is obtained at all. On the other hand, the probability to obtain a certain outcome depends on $\vartheta$, however the outcome is random, and can only guess about the initial state. Only the state orthogonal to the measurement result can be excluded - all other states are in principle possible.

However, it is possible to determine an unknown state with arbitrary precision if an ensemble of $N$ identically prepared copies of the unknown quantum state $|\psi\rangle$ is available. If we perform a $z$-measurement on each of the $N$ independent copies, we obtain a sequence of $N$ independent measurement results corresponding to the probability distribution $\{p_0,p_1\}$. From the statistics of obtained measurement results, one can then determine the parameter $\vartheta$. To this aim, one calculates the relative frequency of measurement results,
\be
\frac{N_0 (+1) + N_1 (-1)}{N},
\ee
where measurement results are specified by the eigenvalues $+1$ (which was obtained $N_0$ times) and $-1$ (which was obtained $N_1$ times), and $N=N_0+N_1$. For large $N$, the relative frequency approaches the expectation value,
\be
\label{ExpSz}
\langle \sigma_z \rangle_{|\psi\rangle} =p_0 (+1) + p_1 (-1) = \cos^2\frac{\vartheta}{2} - \sin^2\frac{\vartheta}{2}= \cos \vartheta,
\ee
where the variance and thus the measurement precision scales as $1/\sqrt{N}$. By choosing $N$ large enough, one can determine $\vartheta$ with desired (arbitrary) precision.

To obtain information about the parameter $\varphi$, measurements on additional copies in a different basis, e.g. $x$-basis, are required. The probability to find the state $|0_x\rangle$ (corresponding to measurement result $+1$) when performing an $x$-measurement is given by
\be
p_0=|\langle \psi|0_x\rangle|^2=\frac{1}{2}|\cos\frac{\vartheta}{2} + \sin\frac{\vartheta}{2}e^{i \varphi}|^2.
\ee
It follows that the expectation value of the observable $\sigma_x$ is given by
\be
\label{ExpSx}
\langle \sigma_x\rangle_{|\psi\rangle}=p_0(+1)+p_1(-1)= \cos\varphi \sin\vartheta.
\ee
If the value of $\vartheta$ is already known from previous measurements, one can then determine the value of $\varphi$ from the relative frequencies of $x$-measurements. Note, however, that $\varphi$ is not uniquely determined (recall that $\varphi \in [0, 2\pi)$, where $\sin$ is not bijective). To completely determine $\varphi$ and hence the quantum state, an additional measurement sequence is required, e.g. $y$-measurements.

If we consider the problem to determine an unknown quantum state on the Bloch sphere representation, one needs to figure out the orientation of the state vector in space. To this aim, it is sufficient to determine the projections of the vector onto the $x$,$y$ and $z$-axes. This corresponds exactly to the expectation values of the observables $\sigma_x$, $\sigma_y$ and $\sigma_z$, respectively. This can be verified by calculating the expressions for the expectation values (see Eqs. (\ref{ExpSx}) and (\ref{ExpSz})), and comparing them with the length of the projections of the Bloch vector, i.e. the absolute values squared of the spherical coordinates.

\begin{figure}[ht]
\centering
\includegraphics[width=7.5cm]{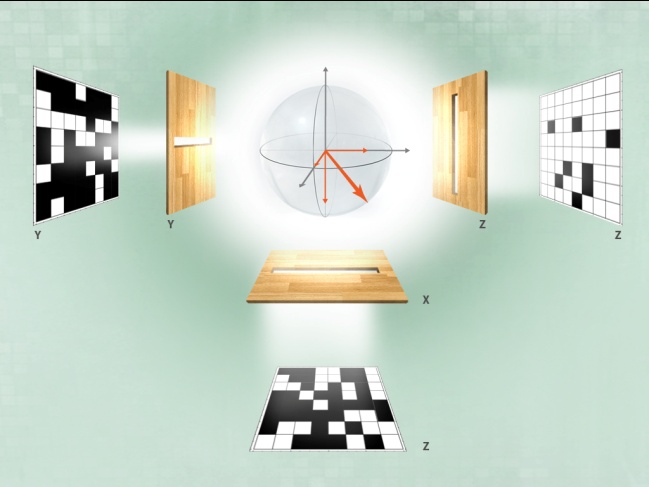}
\caption{State tomography: determining an unknown quantum mechanical state requires a large number of measurements performed on an ensemble of identically prepared qubits. From measurements in $x-$, $y-$ and $z-$ direction, the expectation vaules of the observables $\sigma_x$, $\sigma_y$ and $\sigma_z$ can be determined, which corresponds to the projections of the Bloch vector onto the $x$-,$y-$ and $z-$ axes, respectively.}
\label{Fig16}
\end{figure}

In Fig. \ref{Fig16} the complete measurement process for quantum state tomography is illustrated. The orientation of the slit in $z$-direction corresponds to the measurement of the observable $\sigma_z$. That is, if we observe the state vector, our view is restricted to a single dimension - for example, the $z$-axes.  The measurement result $+1$ is visualized by $\blacksquare$, the result $-1$ by $\square$. By counting the number of black and white boxes, the projection of the Bloch vector on the $z$-axes can be determined experimentally. Similar results hold for observables $\sigma_x$ and $\sigma_y$, where the slit (and hence our view on the system) is restricted to the $x$- or $y$-axis, respectively.

%--------------------------------------------------------------------------------------------------------------------

\subsection{No-cloning theorem}
From the principles discussed in the previous section, it follows that quantum information cannot be cloned or copied. This result is known as "No-cloning theorem" \cite{Wo82}. As shown in the previous section, quantum information cannot be simply read out, preventing the possibility to copy the state of a single quantum system (notice that this is in fact the way copying works in the classical case). To determine the state of a qubit (see Eq. \ref{EqStateBloch}), one needs to identify the value of the parameters $\vartheta,\varphi$ which are real numbers specified by (infinitely) many bits. From a single copy, however, at most one bit of information can be determined, and hence several identical copies are required.

\begin{figure}[ht]
\centering
\includegraphics[width=7.5cm]{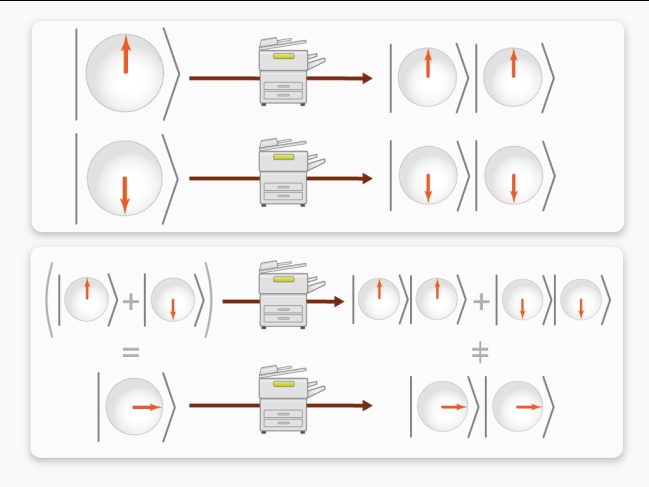}
\caption{Illustration of the No-cloning theorem. Any copy machine capable of successfully cloning states $|0\rangle$ and $|1\rangle$ would provide an incorrect result for the superposition state $|0\rangle +|1\rangle$. Hence no perfect quantum cloning machine can exist.}
\label{Fig17}
\end{figure}

Assume that we can built a cloning machine that successfully copies the states $|0\rangle$ and $|1\rangle$. However, with this the action of the (unitary) process is already fixed:
\bea
|00\rangle \rightarrow |00\rangle &,& |10\rangle \rightarrow |11\rangle,
\eea
and one can easily show that such a machine cannot successfully copy a superposition state $|\psi\rangle=(|0\rangle + |1\rangle)/\sqrt{2}$,(see Fig. \ref{Fig17}). Mathematically, this
follows from
\be
|\psi\rangle|0\rangle \rightarrow (|00\rangle + |11\rangle)/\sqrt{2} \not= |\psi\rangle|\psi\rangle.
\ee
One can in fact show in full generality, that there is no quantum mechanical process that allows to clone the state of an unknown qubit. Only the production of two (or more) imperfect copies is possible, where quantum information is distributed among several qubits. The no-cloning theorem reflects a central feature of the quantum world, which also leads to interesting applications, e.g. in the context of quantum cryptography.

%--------------------------------------------------------------------------------------------------------------------

%--------------------------------------------------------------------------------------------------------------------

\subsection{Sequences of measurements}
\begin{figure}[ht]
\centering
\includegraphics[width=7.5cm]{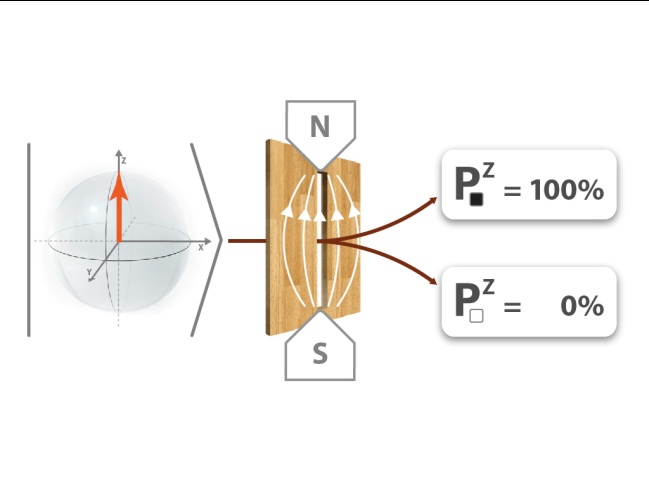}
\caption{A filter can be realized by blocking one branch of a Stern-Gerlach measurement device. Blocking e.g. the upper branch, only particles in state $|1\rangle$ after the measurements can pass. A qubit prepared in $|0\rangle$ will never pass the filter. }
\label{Fig18}
\end{figure}

In the following, we will consider sequences of different measurements and the corresponding probabilities. Consider a system in state $|0\rangle$, where a $z$-measurement is performed. The measurement result will always be $+1$ (corresponding to $|0\rangle$), i.e. the probability to obtain the outcome $-1$ (corresponding to $|1\rangle$) is zero. If one modifies the measurement apparatus in such a way that it acts as a filter where only qubits in state $|1\rangle$ can pass, then all qubits will be blocked. Note that in case of a Stern-Gerlach apparatus, this can be easily achieved by blocking one of the output pathes -- and similarly in the case of a polarization measurement for photons with help of a polarizing beam splitter (see Fig. \ref{Fig18} and Sec. \ref{Photon},\ref{Spin}).

\begin{figure}[ht]
\centering
\includegraphics[width=7.5cm]{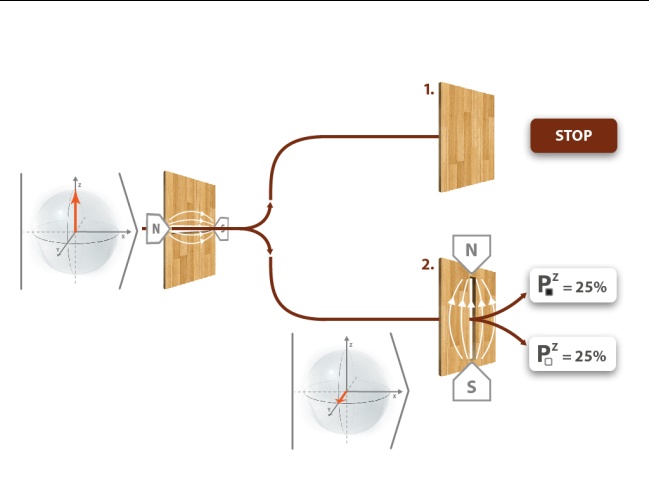}
\caption{Sequence of measurements in different measurement bases. Adding an additional filter oriented in $x$-direction (i.e. only states $|1_x\rangle$ can pass) before the $z$-filter described in Fig. \ref{Fig18}, a qubit can pass the two filters with probability $1/4$, even though it can not pass without the additional filter.}
\label{Fig19}
\end{figure}

Now, we consider a second filter that corresponds to a $x$-measurement (measurement of the observable $\sigma_x$), where quibits with the property $|1_x\rangle =(|0\rangle -|1\rangle)/\sqrt{2}$ (corresponding to measurement result $-1$) can pass. If such a filter is placed before the $z$-filter, something puzzling occurs (see Fig. \ref{Fig19}):
The initial state $|1\rangle$ will pass with probability $p_0=1/2$ through the first $x$-filter, as the measurement yields the result $-1$ corresponding to the state $|1_x\rangle=(|0\rangle -|1\rangle)/\sqrt{2}$ with probability $1/2$. The state of the system has however been altered by the measurement, thus, the input state for the second filter is now given by $|1_x\rangle$. Consequently, this state will now pass through the $z$-filter with probability $1/2$, as the $z$-measurement on $|1_x\rangle$ gives the outcome $-1$ with probability $1/2$. In total, the probability that the particle passes both filters is given by $1/4$, i.e. on average, one out of four particles passes both filters. In contrast, if only the $z$-filter is present, no particle passes! Due to the presence of a second filter, the probability that a particle passes is increased.

An analogous experiment for light can be performed in class using the following simple set-up of polarisation filters: no light can pass through two orthogonal polarization filters, however if an additional filter (oriented $45^\circ$) is placed between the two filters, light passes and the outcoming beam has $1/4$ of the intensity of the incoming beam. Note that only if the experiment is performed with single photons, one can talk about a quantum effect. At this level, it is a purely classical wave effect (see also discussion in Sec. \ref{Photon}). However, also here the interaction with the filter (i.e. the "measurement") change the state of the system.

%--------------------------------------------------------------------------------------------------------------------

\subsection{Heisenbergs uncertainty relation}
We will now consider measurements of different properties or observables of a single qubit. We take into account that measurements in different bases are ''complementary'', as these measurements refer to different properties of the system. One has to decide which of the complementary properties is measured -- only one of the properties can be determined, a measurement of both properties simultaneously is not possible. Or in other words: one cannot say that the system possess both properties simultaneously.
Consider for instance a qubit in the state $|0_x\rangle=(|0\rangle +|1\rangle)/\sqrt{2}$. A measurement of the observable $\sigma_x$ yields a deterministic result, $+1$ and the property $|0_x\rangle$ ($+1$ in $x$-direction) is completely determined. A measurement of the observable $\sigma_z$ in contrast yields a completely random outcome - the state $|0_x\rangle$ does not possess the property $|0\rangle$ or $|1\rangle$ - this property is undetermined and cannot be associated with the system. That is, a qubit cannot have the properties $|0\rangle$ and $|0_x\rangle$ at the same time. On the Bloch sphere picture, this is expressed by the fact that if the state vector points in a certain direction (e.g. along the positive $x$-axes), then the property "pointing up" or "pointing down" are not specified, and similarly for pointing in $\pm y$-direction (corresponding to a $\sigma_y$ measurement). Notice that after the $z$-measurement, the state of the qubit has however the (fixed) property of pointing in $+1$ or $-1$ $z$-direction, as the measurement gives a random but fixed outcome, and the state has changed. However, then the property of pointing in $+x$ direction is no longer present.

Formally, this is reflected in the expectation value and variance of the state for the different measurements:
\bea
\langle\sigma_x\rangle_{|0_x\rangle}= 1 &;&  V(\sigma_x)_{|0_x\rangle} = 0, \nonumber\\
\langle\sigma_z\rangle_{|0_x\rangle}= 0 &;&  V(\sigma_z)_{|0_x\rangle} = 1.
\eea
We have used that
\bea
V(\sigma_{\bm a})_{|\psi\rangle}&=&\langle(\sigma_{\bm a} - \langle \sigma_{\bm a}\rangle)^2 \rangle_{|\psi\rangle} \nonumber\\
&=& \langle \sigma_{\bm a}^2\rangle_{|\psi\rangle} - \langle \sigma_{\bm a}\rangle^2_{|\psi\rangle} \nonumber\\
&=& 1-\langle \sigma_{\bm a}\rangle^2_{|\psi\rangle}
\eea
The variance measures the fluctuations around the expectation value. No variance, $V=0$, means that there are no fluctuations and one can assign a fixed value to the corresponding property. In turn, if the variance is large, no fixed value can be assigned, and the observable is undetermined prior to the measurement. This now implies that if one can assign a fixed (''sharp'') value to one of the observables (i.e. the variance is 0 and the property with respect to this observable is fixed), the variance of the second (complementary) observable is large, and its value is not determined. For the observables $\sigma_x$ and $\sigma_z$ this is not only true for the specific state $|0_x\rangle$, but for any state $|\psi\rangle$: there exists no state $|\psi\rangle$ such that the variance with respect to both observables is small. In this sense, the observables $\sigma_x$ and $\sigma_z$ (and also $\sigma_y$) are complementary.

This corresponds to the same situation as in the famous Heisenberg uncertainty relation for the variances of position and momentum, which can be quantified by $\Delta x \Delta p \geq \hbar/2$, where $\Delta x= \sqrt{V(x)}$. Also in this case, position and momentum refer to different properties of the quantum system - or equivalently the measurements correspond to measurements in different bases, the position basis and the momentum basis. Position basis and momentum basis are related via a Fourier transform, and Heisenberg's uncertainty principle in this case only reflects the fact that functions that are well localized in position space, are not localized in momentum (or frequency) space, leading to large fluctuations or a large variance upon measurement. It is hence not possible to assign fixed values of position and momentum to a quantum system simultaneously. It is also interesting to note that $z$-basis and $x$-basis of a single qubit are also related by a (discrete) Fourier transformation, given by $U=|0\rangle\langle0_x| + |1\rangle\langle 1_x|$ -- the so-called Hadamard operation. Notice that the uncertainty relation also holds for other complementary variables, where the above mentioned explanation using Fourier transform is not applicable.

%--------------------------------------------------------------------------------------------------------------------

\subsection{Random number generators}
The generation of true randomness, i.e. sequences of (truly) random numbers- is of fundamental importance for many technical and scientific applications, e.g. for numerical simulations. In many cases, so-called pseudo random number generators are used, which produce random numbers following certain algorithms. This is, however, not sufficient for many applications. The measurement of a single qubit offers the possibility to obtain true random bits and hence true random numbers. To this aim, a qubit is prepared in the state $|0_x\rangle$ and measured in the $z$-basis (observable $\sigma_z$). This yields a completely random outcome. Such quantum random number generators can be commercially purchased \cite{Industry}.

%--------------------------------------------------------------------------------------------------------------------

\subsection{Quantum cryptography - BB84 protocol}
In this section, we will discuss a modern practical application of single qubits in the context of quantum cryptography \cite{QCrypto}. The aim is the secure transmission of secret messages from a sender (Alice) to a receiver (Bob). The transmission of single qubits across a quantum channel is used to establish a secret key ---a sequence of random bits that are only known to Alice and Bob, but not to any adversary. Using these random bits, a message of the same length as the secret key can be securely transmitted using the so-called one time pad. The random bits are first added to the message bits (modulo 2) at the sender, and after transmission the same random bits are subtracted by the receiver. Adding random numbers washes out any structure of the initial message, as the resulting bit sequence is still completely random. Hence decryption schemes that makes use of the structure in the message fail, and in fact it can be shown that such a transmission is provably secure if the parties share a random key of the same length as the message.

To establish such a secret key, several schemes have been proposed whose security is based on physical principles of quantum mechanics. We will discuss the BB84 protocol, named after its inventors C. Bennett and G. Brassard. The security of the protocol is based on the fact that any attempt to reveal information about a transmitted quantum system involves a measurement, and hence leads to change of the transmitted state (as long as the measurement direction does not coincide with the state of the qubit). This change of the state can be detected and allows one to detect the presence of an eavesdropper. In addition, it is impossible to clone the state of a single qubit, or to determine its unknown state if only a single copy is available. These aspects can be used to design a protocol where the security of the scheme is guaranteed by the laws of nature ---more precisely by the behavior of quantum systems under measurements. This is in stark contrast to classical cryptographic schemes such as the widely used RSA-encryption, whose security are based on (unproven) assumptions on the complexity or difficulty to compute certain functions or perform certain tasks, e.g. calculating the prime factors of a large number (factoring problem). Notice that precisely this problem can be solved efficiently with help of a quantum computer, rendering classical cryptographic systems insecure once a quantum computer could be build.

In the BB84 protocol single qubits are transmitted from Alice to Bob, and subsequently measured. To this aim, Alice (randomly) selects a bit value $j \in \{0,1\}$, and also randomly a basis $\alpha \in \{x,z\}$. She prepares the qubit in the state $|j_\alpha\rangle$, i.e. one of the four states $\{|0\rangle,|1\rangle,|0_x\rangle,|1_x\rangle\}$. Bob randomly selects a measurement basis $\beta \in \{x,z\}$ and performs a measurement in the $\beta$-basis (observable $\sigma_\beta$) on the received qubit. If the measurement basis coincides with the preparation basis, i.e. $\alpha = \beta$, Bob obtains a deterministic outcome, $|j_\beta\rangle$, from which he can determine the bit value $j$. For $\beta \not= \alpha$, the measurement result is random. For example, if $\alpha=x$ and $j=0$, the state $|0_x\rangle$ is transmitted, and a $z$-measurement ($\beta$=z) leads to outcome $|0\rangle$ and $|1\rangle$ with probability $p_0=p_1=1/2$.

Now a total of $N$ random qubits are transmitted to Bob, who measures each of the qubits. In each round, the bit value $j$ and the preparation bases $\alpha$ are randomly and independently selected by Alice, and also Bob randomly selects the measurement basis $\beta$ in each round. Afterwards, the used bases $\alpha$ and $\beta$ are revealed through a public channel. That is, this information is available to an eavesdropper. However, only the preparation and measurement bases are revealed - not the prepared or measured bit value $j$. If $\alpha=\beta$ in a specific round, than we know from above considerations that Bobs measurement result coincides with the prepared state, i.e. Alice and Bob share a random bit $j$. That is, after $N$ rounds Alice and Bob share a sequence of approximately $\tilde N \approx N/2$ random bits. As a final check, Alice and Bob randomly select $M$ of theses bits and compares the bit values over the public channel. If all of these $M$ bits coincide, Alice and Bob can deduce that an eavesdropper can only have gained an (exponentially) small amount of information on their sequence of random bits. The remaining $\tilde N - M$ bits can then be used as a random bit string, i.e. as a key for cryptography and hence to securely transmit secret information. If many of the selected bits do not coincide, one has to assume that an eavesdropper has interfered with the transmission, and has potentially learned information about the whole bit string. In this case, the protocol is aborted and no message is transmitted. Notice that only the generation of the key failed, and the eavesdropper does not possess parts of the message.

We now discuss the security of the protocol. To this aim, we consider different eavesdropping strategies.

The first (extremal) strategy for the eavesdropper Eve is to do nothing and simply guess the bit value. For each bit, the correct value is guessed with probability $1/2$, while Alice and Bob clearly do not detect anything. However, the probability to correctly guess the correct value of the whole bit string is given by $p=(1/2)^{\tilde N}$, i.e. is exponentially small.

Another possible strategy is that Eve stores all transmitted qubits until Alice and Bob reveal the used preparation basis, and simply send different qubits in a (randomly) selected state to Bob. In this case, Eve learns the bit value of each qubit, and hence the whole bit string. However, when Bob performs a measurement on the received qubit, the bit value he obtains will only with probability $p=1/2$ correspond to the transmitted value by Alice. That is, in the control step, many of the checked bits will not coincide. In fact, the probability that all of them coincide is given by $p=(1/2)^M$, and hence Eve will be detected with probability $1-(1/2)^M$.

An intermediate strategy consists of performing a measurement in a randomly selected basis $\gamma \in \{x,z\}$ on each of the qubits once it is transmitted - this is called an individual attack. The method resembles the scenario described in Fig. \ref{Fig19}. If Eve guesses correctly, i.e. $\gamma = \alpha$, she learns the bit value and the measurement does not change the transmitted state. Hence in this case, Eve will also not be detected in the control step. However, if Eve measures in the wrong basis, i.e. $\gamma \not= \alpha$ (which happens with probability $1/2$), the transmitted state is changed. For example, if $|0_x\rangle$ was transmitted, and Eve measures in the $z$-basis, the state after the measurement is given by $|0\rangle$ or $|1\rangle$. In both cases, Bob will obtain a random outcome if he performs an $x$-measurement. That is, with probability $1/2$ he will obtain the correct bit value, however with probability $1/2$ he will obtain the wrong value - which will subsequently be detected in the control step. In turn, since Eves measurement result is random, she will guess the correct bit value with probability $1/2$. In total, if Eve uses this strategy, she will learn the correct bit value with probability $p=3/4$, while Bob will obtain the wrong bit value with probability $p_{\rm error}=1/4$ (notice that both branches, $\gamma=\alpha$ and $\gamma\not=\alpha$ occur with probability $1/2$). Thus Eve knows the correct bit string with probability $(3/4)^{\tilde N}$, while the probability that an error is detected in the control step is given by
\be
1-(1-p_{\rm error})^M= 1-\left(\frac{3}{4}\right)^M.
\ee
If $M$ is sufficiently large, the eavesdropping attempt will be detected with almost certainty.

Notice that the security cannot only be shown for above mentioned strategies, but in fact for {\em any} strategy, including strategies that involve the storage and joint measurement of all qubits. Any information gain of Eve leads to a change of the transmitted states, i.e. an error in bit value obtained by Bob. This error can then be exponentially amplified and hence detected in the control step by checking a sufficiently large number $M$ of bits. In turn, one can conclude that if the number of errors is sufficiently small, (almost) no information of the overall bit string is known to Eve. A secure transmission of the message is hence possible.

One should stress that we have considered a somewhat idealized scenario here. On the one hand, one assumes perfect quantum channels, which in reality will however be noisy, and errors induced by channel noise are indistinguishable from possible eavesdropping attempts. There are however strategies ---e.g. making use of classical privacy amplification---, which allow one to deal with such situations if the channel noise is sufficiently small ($\approx 10\%$). On the other hand, we have assumed in the analysis that we are dealing with perfect qubits and perfect measurement devices. In experimental realizations, this is however not the case: laser diodes do not produce single photon states, but with a certain probability also multi-photon states following a Poissonian distribution. This can be used for so-called Trojan-horse attacks, where the additional degrees of freedom are used to undermine security. Furthermore, the current technological implementation of single photon detectors allows for eavesdropping strategies, where the detectors are effectively controlled from the outside by strong light beams \cite{QCryptoExp}. These technological problems can be solved in principle. From a theoretical point of view, new strategies are developed that are known under the name of  ''device independent quantum cryptography'' which aim at minimizing the requirements for the technological implementation. Its worth mentioning that quantum-cryptography systems based on the BB84 protocol are already commercially available and under use in financial industry \cite{Industry}.

%--------------------------------------------------------------------------------------------------------------------
%--------------------------------------------------------------------------------------------------------------------
\section{Summary and conclusions}
%--------------------------------------------------------------------------------------------------------------------
%--------------------------------------------------------------------------------------------------------------------
In this article, we discussed the simplest quantum mechanical system -the qubit- in full detail. As shown in the text, many of the central concepts and principles of quantum mechanics can be illustrated using a single qubit. This approach involves no complex mathematical descriptions or advanced mathematics --e.g. wavefunctions and Schr\"odinger equation, Hilbert spaces etc. It suffices to consider simple vectors for a quantitative description. Perhaps more importantly, we put forward a qualitative description based on simple visualizations on the Bloch sphere. With this approach, not only the superposition principle, but also the strange behavior of quantum systems under measurements can be described and illustrated. Furthermore, the differences to classical physics can be highlighted.

We do not restrict ourselves to a single, specific physical realization of a qubit. In our approach, we introduce the qubit as an abstract object, and only later discuss different physical realizations. We have discussed advantages and potential problems which might occur in class, when the qubit is introduced as abstract object on the Bloch sphere. A careful analysis of the features of a qubit does not only allows for a discussion of fundamental concepts such as Heisenbergs uncertainty principle, but also for the treatment of modern (technological) applications in quantum information theory and quantum communication. We hope that we could illustrate with our approach the possibility to make students familiar with some of the basic principles of one of the most important --but also most counter-intuitive-- physical theories to date - quantum mechanics, avoiding advanced mathematics or mythical jargon.

%--------------------------------------------------------------------------------------------------------------------

\section*{Acknowledgements:} We thank Dipl. Des. Michael Tewiele for the pictures.

{\em Note added: A german version of this article was published in \cite{Du12}.}

\end{document}